\documentclass[aip, %
  amsmath,amssymb,%
  reprint,%
]{revtex4-1}

\usepackage[T1]{fontenc} % font
\usepackage[utf8]{inputenc} % font
\usepackage{lmodern} % font
\usepackage{microtype} % better justified text
\usepackage{eurosym}% \euro
\usepackage{physics} % for physics math commands
\usepackage[range-phrase = --,retain-unity-mantissa = false,exponent-product = \cdot]{siunitx} % dimensioned quantities
\usepackage{graphicx} % Required for inserting images
\graphicspath{{},{./}}
\usepackage{float} % force placement with H

\DeclareSIUnit{\pixel}{px}
\DeclareSIUnit{\px}{px}
\DeclareSIUnit{\frame}{frame}
\DeclareSIUnit\litre{l} % The war between L and l rages...

\newlength{\figwidth}
\renewcommand{\emph}[1]{\textsl{#1}} % emphasis (lighter italic)

\usepackage[normalem]{ulem} % pour barrer des trucs
\usepackage{todonotes} % \todo{foo}
\usepackage{soul}
\sethlcolor{orange}
\makeatletter
\if@todonotes@disabled

\else

\fi
\makeatother
\newcommand{\Ca}{\ensuremath{\mathrm{Ca}}}
\newcommand{\Bo}{\ensuremath{\mathrm{Bo}}}
\newcommand{\fr}{\vb{q}} % q or Q
\newcommand{\hcrest}{{h_\text{crest}}} % h_crest or h_top or h_t or h_c
\newcommand{\ellc}{{\ell_\text{c}}} % l_c or l_cap
\newcommand{\hnorm}{h^*} % h^* or \tilde{h}

\begin{document}

\title{Controlling deposition and characterising dynamics of thin liquid films with high temporal and spatial resolution}
\author{G. Le Lay}
\email{gregoire.le-lay@u-paris.fr}
\author{A. Daerr}
\affiliation{Matière et Systèmes Complexes UMR\,7057 Université Paris Cité, CNRS,
  75231 Paris cedex 13, France}%

\date{\today}% It is always \today, today,

\begin{abstract}
  The high inertia of classical fluid coating processes severely
  limits the possibility of controlling the deposited film thickness
  through the entrainment velocity.
  We describe and characterize a new experimental device
  where the inertia is dramatically reduced, allowing for
  millimeter-scale patterning with micrometer-accurate thickness.
  Measuring precise film profiles over large spatial extents with high
  temporal resolution poses a challenge, which we overcome using a
  custom interferometric set-up coupled with state-of-the-art
  signal processing.
  The sensitivity of our method allows us
  to resolve film thinning rates in the
  nanometer-per-second range, and to quantify the relative
  contribution of surface-tension and gravity driven flows.
  We apply this method by showing that the thickness of the deposited
  film obeys the classical Landau-Levich
scaling even when the
  meniscus faces important acceleration.

\end{abstract}

\maketitle

Fluid entrainment by a moving substrate is of interest in many contexts,
notably as a process to coat solids with thin layers.
From wire insulation to the modification of optical and surface properties of window glass,
controlling the thickness and uniformity of fluid layers is of key importance.
The possibility to create textures is also increasingly explored,
e.g. by controlling instabilities arising in the presence of evaporation\cite{deegan1997,berteloot2012}.
In confined flows,
the entrainment transition governs the pressure head required to drive multi-phase flows such as Taylor bubbles,
because of the high viscous dissipation in the displaced menisci.
This is of importance because these flow arise in many different applications,
including the study and design of microfluidic devices\cite{angeli2008},
or the prevention of airway obstruction in biomedical context\cite{grotberg2011, baudoin2013}.
Last, multi-phase flows are of particular interest in the study of vapour–liquid mass exchange in heat pipes\cite{zhang2021}.
In this domain, thin films have gained attention as an efficient way to improve
transport in capillary driven heat exchangers\cite{zhang1998, nikolayev2021},
and require understanding its evolution in non-stationary systems such as loop heat pipes\cite{launay2007}.

The thickness of an entrained film has been studied in various
gemetries such as flat plates\cite{goucher1922,morey1940,lasseux1991,quere1997},
capillaries\cite{davies1950, bretherton1961}, or fibers\cite{quere1999}. Its
asymptotic scaling was derived by Landau, Levich and Derjaguin (LLD)
\cite{landau1942,derjaguin1943,white1965}, who showed that the
steady-state film thickness of the fluid layer of viscosity $\eta$ and
surface tension $\gamma$ entrained by a planar solid pulled from a bath
results from the compensation of viscous stresses by capillary
pressure gradients at the bulk-film transition.
Limits and corrections to this famous result have been extensively
explored\cite{ruschak1985,quere1999}, such as effects of gravity,
evaporation, Marangoni forces in the presence of
surfactants\cite{shen2002}, wetting dynamics near contact
lines\cite{teletzke1988,snoeijer2006}, non-Newtonian rheology/suspensions, or inertial and boundary layer effects
at high speeds\cite{deryck1996inertial}.

In order to visualise and quantitatively measure the features of the entrained films,
several techniques are used in the literature.
Direct optical visualization of the system only allows to see the meniscus, which refracts light strongly,
but prevents measurements on the film itself\cite{lips2010}
The thickness of a very thin film can be precisely measured at one point,
either by spectrometry of the reflected signal\cite{snoeijer2008, fourgeaud2017},
or by measuring the displacement of the focus points of a laser\cite{youn2018}.
The point probe can then be displaced, if the film evolution is periodic.
The deflection of a grid by the interface can be used to reconstruct the film profile\cite{fourgeaud2017},
with a spatial resolution limited by the grid spacing.
The visualization of one-directional interference patterns for different wavelengths\cite{zheng2002} also allows to measure the film profile.
Often, several of these techniques are used in combination in order to cumulate their advantages.

The purpose of this paper is to present an original set-up to deposit
and characterize films of controllable thickness through the
displacement of a fluid rivulet in a Hele-Shaw geometry.

We explore the potential of this system for non-uniform, high-dynamic
range coating, and show that the film can be modulated at sub-micron
scale in thickness and at sub-millimetric scale in the direction of
meniscus motion. This is possible because the moving fluid phase has a
very small volume, so that its inertia is vastly reduced almost to
that of the surrounding gas phase, in stark contrast to the bulk
displacement of one fluid by another as studied by Saffman, Taylor and
others\cite{saffmantaylor1958, tabeling1987}, or to moving substrate
configurations. Fast acceleration can therefore be achieved by
moderate pressure jumps.

We monitor the film evolution using an interferometric measurement
technique to study the structure and dynamics of the deposited film
with both high spatial and temporal resolution. The precision of the
obtained profiles is such that we can quantify the contribution of
surface-tension induced flows that depend on the fourth order spatial
derivative, and compare it to the gravitational drainage. We are also
able to verify that the Landau-Levich-Dejarguin (LLD) scaling for the
film thickness holds even when the meniscus accelerates from static to
maximum velocity on a length scale roughly equal to the outer meniscus
scale.

After describing our experimental set-up (\ref{sec:setup}) and
measurement method (\ref{sec:spacetime}), the investigation of the
influence of capillary flow versus gravitational drainage is
investigated in section~\ref{sec:drainage}, and the relation of film
height to meniscus speed for a sinusoidally driven liquid bridge is
studied in section~\ref{sec:periodicdeposition}.

\section{Film deposition and observation} \label{sec:setup}

To generate a very thin layer of liquid of controlled height on a substrate, we use the displacement of a liquid stream in an air-filled Hele-Shaw.
Between two glass plates (width \SI{10}{\centi\meter}, length \SI{1}{\meter}) separated by a small gap $b$, we flow a continuous steam of liquid so that it wets both plates, forming a liquid bridge between them in which the fluid flows downwards under the action of gravity.
The liquid stream consisting in this vertically-extended bridge laterally delimited by two semicircular menisci is henceforth termed rivulet.
We design by vertical, or streamwise, the $x$ direction, oriented with gravity.
The $z$ direction in which the rivulet can move is termed transverse, and $y$ corresponds to the gapwise direction, being perpendicular to the plates.

The liquid we use in this article is a perfluorinated oil (density $\rho=\SI{1.72}{\gram/\centi\meter\cubed}$, kinematic viscosity $\nu = \SI{1.00}{cSt}$, surface tension $\gamma = \SI{14}{\milli\newton/\meter}$, refractive index $n=1.26$) and we set the gap between the plates to be $b = \SI{0.6}{\milli\metre}$.
We inject the liquid using a gear pump through silicon tubing.
The rivulet falls down vertically, and this situation is stable while we stay under a critical flow rate \cite{daerr2011}.
However, the rivulet can be locally displaced by imposing a pressure difference between the right and left sides of the cell.
Indeed, since the rivulet separates the cell vertically in two parts and since it is airtight, it behaves like a ``liquid membrane'' and can be pushed transversally using an imposed pressure difference between both sides --- or, equivalently, by pushing in with an asymmetric air flow.
In our experiments we use speakers on both side of the cell to control the lateral movement of the rivulet.
By feeding these speakers opposite signals, when the membrane of one advances, the membrane of the other retracts, pushing the rivulet in the same direction \cite{lelay2025}.
Note that by using static pieces in the cell, we can change the geometry of the air flow generated by the speakers and locally modulate the amplitude of movement of the rivulet.

\begin{figure}%[H]
\centering
\includegraphics[width=\figwidth]{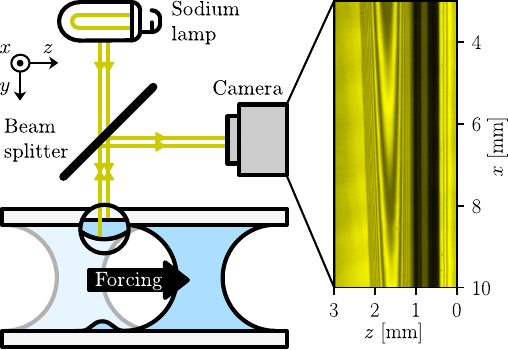}
\caption{Visualization setup.
The rivulet is pushed, leaving a thin liquid film behind it.
The interference pattern due to the light being refracted on both interfaces are observed by a camera.}
\label{fig:schema_interferences}
\end{figure}

While moving, the menisci deposit behind them very thin liquid films (see fig~\ref{fig:schema_interferences}), which are the main object of study of this paper.
By imposing the signal sent to the speakers, the movement of the rivulet can be precisely controlled.
This allows for deterministic and reproducible film deposition on the plates.

To measure with precision the geometry of such a very thin film (no thicker than \SI{10}{\micro\meter}) as well as its temporal evolution is an experimental challenge.
To adopt an unambiguous convention, we henceforth call height and note $h(x, z, t)$ the spatial extent in the $y$ direction (which also corresponds to the depth, or thickness) of the film.
The films have extremely high aspect ratio, with their height being around a thousand time smaller than their transverse and streamwise extent.
To maximize the resolution in all directions while maintaining a high temporal resolution, use an interferometric method.
As will be shown later, this allows us to measure microscopic changes of film depth over a macroscopic space scale on the other directions, while temporally resolving all phenomena of interest.
We use monochromatic lighting (high-pressure sodium-vapor lamp, $\lambda_0 = \SI{589}{\nano\meter}$) to visualize the film.
The light reflected by the glass-oil and oil-air interfaces creates an interference pattern, that we observe using a beam splitter and a camera as shown on fig. \ref{fig:schema_interferences}.
We use a Manta~G-223B camera controlled by the open-source software Limrendir to record the experimental sequence of images.
We mount a telecentric lens on this camera in order to only visualize the light that arrives normal to the glass surface.
Note that since the menisci limiting the rivulet touch the front and back plates, they leave films on both.
In order to visualize only what happens on one plate, we use a lens with a shallow depth of field, which allows us to focus on the film deposited on the frontmost plate.

The images we obtain consist in successions of bright and dark interference fringes, which corresponds to ``iso-height'' level lines of the film.
Since the topology of the film is encoded in the fringes, it is possible to demodulate the luminosity signal in order to find the height profile of the film.
The luminosity of a fringe depends on the value of the phase difference between the interfering light rays $\Phi = 2\pi\frac{2\,h}{\lambda}$ where $\lambda = \lambda_0/n$ is the wavelength of the light inside the liquid.
Between two fringes of opposite brightness, the height difference corresponds to a phase difference of $\Phi = \pi$.
This corresponds to $h_0 := \lambda/4 \approx \SI{117}{\nano\meter}$.
This means that we are able to measure variations of height corresponding to one-tenth of a micron on spatial scales going from \SI{10}{\micro\meter} (distance between two adjacent pixels) to \SI{10}{\milli\meter} (total field of view of the camera).
Thus, we can monitor the evolution of the geometry of the whole film on arbitrary short (or long) timescales by modifying the frame rate of the camera.

\begin{figure}%[H]
\centering
\includegraphics[width=\figwidth]{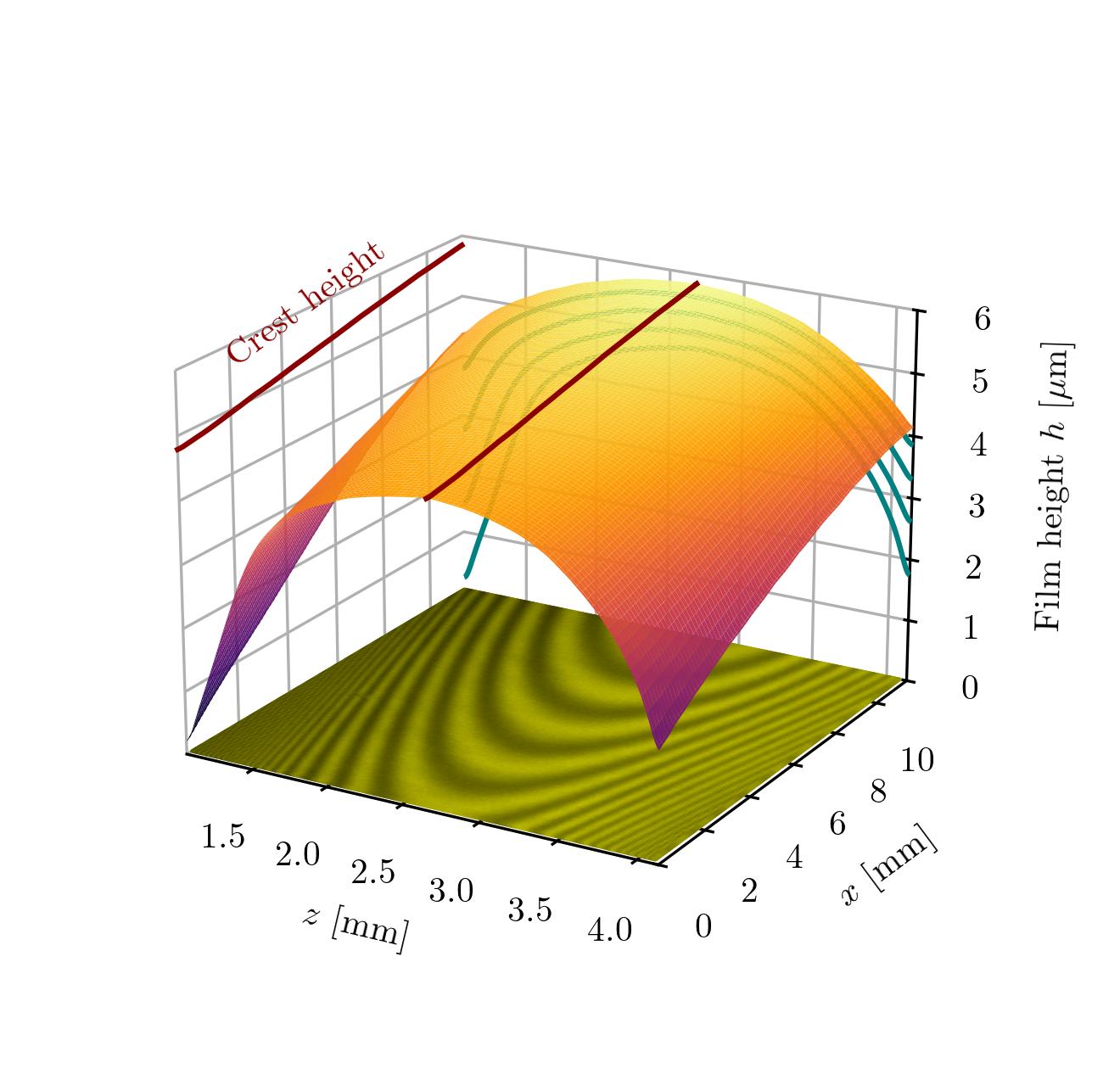}
\caption{Height profile of a liquid film.
The crest is indicated by a dark red line.
The transverse cross-sections are projected to the back, in blue.}
\label{fig:3dtunnel}
\end{figure}

The typical geometry of a film we are interested in is shown on figure~\ref{fig:3dtunnel}.
The changes of luminosity on the image translate to changes of height of the film, that we are able to quantitatively deduce, as shown on the figure.
Note that for the images presented in this article, the color scheme has been chosen in order present a visible contrast between dark and bright interference fringes.
Usually our films have a transverse section ($h(z)$) looking like half of a very flat ellipse (with a high aspect ratio, of order 1000/1), and the height varies very slowly with the $x$ coordinate (with very shallow slopes, of order $10^{-4}$).

In this study, we will be interested in the films generated by two different kind of rivulet movement.

In a first part, we look at films generated by a fast and sudden movement of the rivulet.
This snapping motion leaves behind the rivulet an initially thick film, the evolution of which we can monitor over long timescales.
This allows us to show that precise measurements are possible using our method, and we show that thanks to our high space and time resolution, we are able to quantitatively explain the evolution of the film geometry as a function of time.

In a second part, we are interested in films left behind rivulets adopting a periodic motion.
This allows us to demonstrate that we can create highly reproducible films with a precisely controlled geometry.
\section{Space-temporal evolution} \label{sec:spacetime}
By imposing a snapping motion to the rivulet, we create a film with a relatively sharp transverse profile, due to the rapid change of deposition speed during the movement.
We then monitor the time-evolution of the film, measuring the change in its geometric properties with time.
When relaxing, the geometry of the film changes in two ways: its cross section flattens, and the height of its crest line  decreases due to liquide drainage.
In this paper, we call crest the curve along the $x$ direction formed by all the maxima along the transverse direction, as illustrated on figure \ref{fig:3dtunnel}.

\begin{figure}%[H]
\centering
\includegraphics[width=\figwidth]{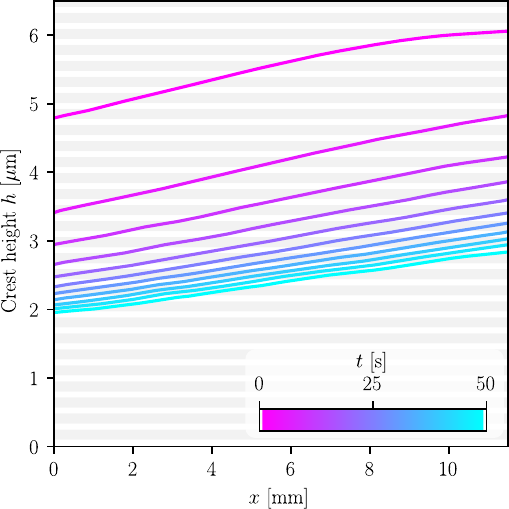}
\caption{Evolution of crest height $\hcrest$ as a function of space (bottom axis) for different times (color/grayscale).
The alternating white and gray horizontal lines have a width (and spacing) of $h_0=\lambda/4$.
They allow to visualize the interference fringes crossed along a crest line.}
\label{fig:drainstudy_heightevolve}
\end{figure}

As is shown on figure~\ref{fig:drainstudy_heightevolve}, we are able to measure the height of the crest $\hcrest$ as a function of space at different times.
As one can notice, the film drains as time goes on, and the height of its crest (and the rest of the film) diminishes with time.
The film also has a slope, due to inhomogeneities of the rivulet movement amplitude.

\begin{figure*}%[H]
\includegraphics[scale=1]{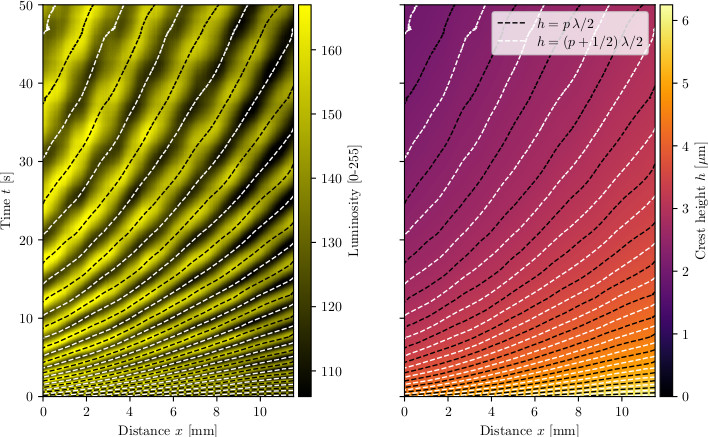}
\caption{Spatio-temporal evolution of the crest height.
Left: reference signal, i.e. normalized light intensity along the crest.
Right: Height profile reconstructed by unwrapping the phase of the left signal.}
\label{fig:drainstudy_2dmap}
\end{figure*}

Thanks to the high time resolution our method allows, we can do even better and push the measurement of the crest height to its limit by visualizing the entire field $\hcrest(x, t)$.
This is done by gathering the light signal along the crest line at every frame with a high frame rate (\SI{50}{\frame/\second}).
The luminosity along the crest as a function of time and space is shown on figure~\ref{fig:drainstudy_2dmap} (left).

Using this information, we can use this information to obtain the phase of the signal.
This is done by using a two-dimensional Hilbert transform, allowing us to recover the phase of the corresponding analytic signal.
Using this instantaneous phase $\Phi(x, t)$ of our signal, we can then reconstruct the entire crest height profile, as is shown on figure~\ref{fig:drainstudy_2dmap} (right).
On this figure, we plotted on both spatio-temporal profiles the level lines corresponding to $h = N\,h_0$, with $N$ an integer being even (black dashed lines) or odd (white dashed lines).
As one can visually confirm, these levels correspond to the bright and dark interference fringes, since they correspond to constructive and destructive interference conditions, respectively, for the reflected light.
We are then able to use this information to quantitatively understand the height evolution mechanism.
\section{Film drainage} \label{sec:drainage}
We have shown that we can experimentally access many difficult to measure metric, such as the height profile $\hcrest$, its slope $\partial_x \hcrest$ and even its evolution with time $\partial_t \hcrest$.
We show in this section that we can quantitatively link these variables, confirming the relevance and the precision of our measurements.

The time-evolution of the height $h(x, z, t)$ of the film is dictated by the thin film equation for a newtonian fluid with no surface shear on a vertical substrate.
It reads
\begin{align} \label{eq:thinfilm}
  \pdv{h}{t} &= -\div{\fr} - \Phi_\text{evap}\qqtext{with} \fr = \frac{h^3}{3\,\rho\,\nu}\qty[\grad\qty(\gamma\laplacian h) + \rho \vb{g}]
\end{align}
where in addition to the already introduced variables $\vb{g} = g\vb{e}_x$ represents gravity, $\fr$ the fluid flux and $\Phi_\text{evap}$ is the evaporative flux.
The fluid flux $\fr$ can be separated into two contributions, due to capillarity and gravity.
Since the film is curved, there is a capillary pressure inside the liquid $\gamma\laplacian h$, the gradient of which is responsible for a capillary flow.
And since the film is deposited on a vertical surface, the volumetric gravity force $\rho \vb{g}$ induces a gravity-driven flow.

When applying this equation to the films that are of interest in our study, we are able to make some simplifications.
First, in our system the films height always vary more intensely in the transverse than in the streamwise direction.
This allows us to neglect spatial derivatives along the vertical direction $\partial_x$ when compared to partial derivatives of similar order along the transverse direction $\partial_z$.
Moreover, since the atmosphere in the channel is only slowly renewed, we can safely consider that the air inside the cell is saturated in liquid in vapor phase.
This is confirmed by the fact that some extremely thin films (shallower than \SI{0.5}{\micro\meter}) present in the field of view do not disappear during the experiment, although they drain no fluid and only evaporate.
Thus, since we do not heat our system, the evaporation $\Phi_\text{evap}$ can be considered negligible in our case.
And since we are interested in the evolution of the crest height $\hcrest(x, t)$, for which $\partial_z h = 0$ (by definition), we obtain
\begin{align} \label{eq:dhdt}
  \pdv{\hcrest}{t} &= -\frac{g}{\nu}\,\frac{\ellc^2\, h^3}{3}\,\eval{\pdv[4]{h}{z}} - \frac{g}{\nu}\,h^2\,\pdv{h}{x} \qqtext{with} h=\hcrest\\
\end{align}
where $\ellc$ is the capillary length $\sqrt{\gamma/(\rho\,g)}$.
The change of height of the film at the crest is the sum of two contributions, which are related to flows of the different physical origins we discussed earlier: capillarity and gravity.
The second term corresponds to advection of the crest profile, characterized by its slope $\partial_x h$, due to the gravity-driven Poiseuille flow.
The first term represents the drainage of fluid towards the edges (or the center) of the film when there is an excess (or a lack) of curvature at the crest.
It depends on the \textsl{snap} $\eval{\partial_{zzzz}h}_\hcrest$ at the crest, which is the geometrical quantity that describes the rate of evolution of the radius of curvature.
This snap term can be written $\partial_{zzzz}h = \epsilon / (R\, L^2)$ where $R$ is the radius of curvature, $L$ is the characteristic length over which the curvature evolves and $\epsilon$ is equal to $-1$ if there is an excess of mass at the center of the film (the transverse profile is sharper than a parabola) and $+1$ if the mass is predominantly on the sides of the film, i.e. the film is ``flatter'' than a parabola.

Whether or not any of these terms dominate can be determined by using a dimensionless number comparing capillarity and gravity, that can be assimilated to the equivalent of a Bond number:
\begin{align} \label{eq:bond}
  \Bo = \frac{3 \, \abs{\partial_x h}}{\ellc^2 \, h \, \abs{\partial_{zzzz}h}} = \abs{\partial_x h} \, \frac{3 R}{h} \, \qty(\frac{L}{\ellc})^2\ .
\end{align}
However, it is difficult to evaluate the order of magnitude of this number.
Indeed, on the one hand, the film is very thin ($h \ll R$), because of the relatively low speeds at which the menisci is driven.
On the other hand, the film is very flat in the vertical direction ($\abs{\partial_x h} \ll 1$), because the amplitude of rivulet movement varies weakly along the $x$ direction.
Hence we can not easily conclude a priori, and we must experimentally measure the geometric properties of the film to determine which of the two effects dominates the film drainage.

\begin{figure}%[H]
  \centering
  \includegraphics[width=\figwidth]{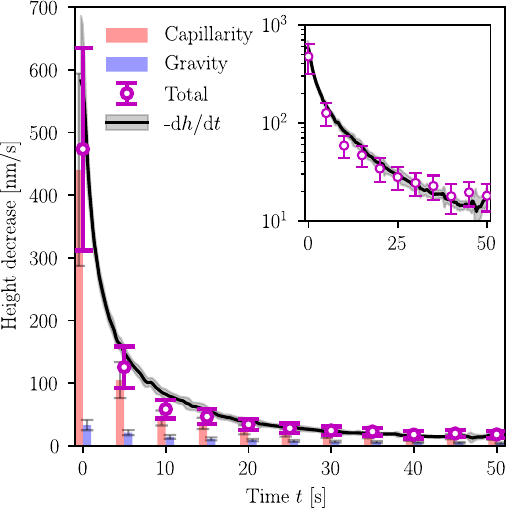}
  \caption{Drainage of the crest of the film as a function of time, with contribution from both capillarity (red) and gravity (blue). The purple points `Total' are the sum of capillary and gravity-driven drainage. The black line corresponds to the direct measurement of $\partial h/\partial t$.}
  \label{fig:drainstudy_conclusion}
\end{figure}

The measure of $\partial_x \hcrest$ is straightforward as it is corresponds to the measure of the spacing between interference fringes along the crest.
The measure of the \textsl{snap} $\partial_{zzzz}h$ however is much more subtle because it requires the measurement of the fourth derivative of an experimental signal.
It is always a challenge to estimate quantitatively and with acceptable accuracy such a high-order derivative of a noisy signal.
Thanks to the high accuracy of the interferometric method, we are able to extract a meaningful measurement of this quantity and its uncertainty, as displayed on supp. fig SF1.
The resulting measurements of the mean value of $h$, $\partial_x h$ and $\partial_{zzzz}h$ can be seen on fig SF2.
From these measurements, we are able to compute the magnitude of both terms on the right-hand side of equation \eqref{eq:dhdt}.
And since our time resolution allows it, we can also compute $\partial_t h$ with a comparable accuracy, in order to verify our predictions.
The comparison between the direct measurement of the height decrease and the computed effects due to capillarity and gravity can be seen on figure~\ref{fig:drainstudy_conclusion}.
As can be seen in this figure, in the case of a snapping motion, the decrease of crest height is mostly imputable to capillarity.

Our technique thus allows us to quantitatively measure geometrical characteristics of thin films, with excellent accuracy both in space and time.
This is showcased by our ability to accurately measure the time evolution of the films, and link it to their physical causes.
\section{Controlled deposition of film} \label{sec:periodicdeposition}
Another interest of our setup is the ability to decide the initial characteristics of the deposited film by controlling the movement imposed on the rivulet.
Indeed, since the films are deposited by the menisci on the edges of the rivulet, pushing it at a controlled speed leads to the deposition of a film of deterministic thickness.
To illustrate this approach, in this section we drive the rivulet back and forth periodically in the transverse direction using speakers on the side of the cell.
By measuring independently the deposition speed and film thickness, we confirm that we can predict the depth of the resulting film.

\begin{figure*}%[H]
  %\centering
%
  \includegraphics[scale=1]{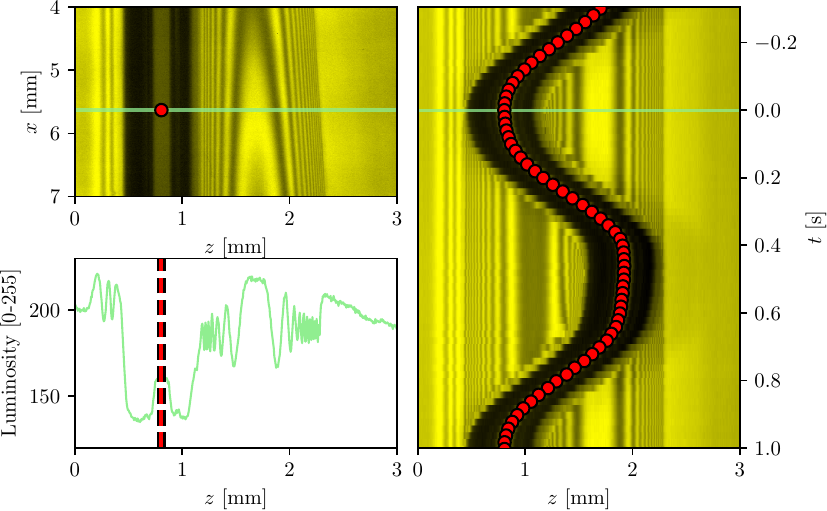}
  \caption{Extraction of the position of the rivulet on the image.
    (a) Experimental image for $t=0$. In green, the position $x_3$ which we consider.
    The black and red dot is the (detected) position of the rivulet.
    (b) Luminosity profile for $x=x_3$ with the rivulet position in black and red dashed line.
    (c) Slice of the image for $x=x_3$ as a function of time, with the position of the rivulet.}
  \label{fig:profile_and_speed}
\end{figure*}

Since the incident light is strongly refracted by the menisci on the
side of the rivulet, we see it in the images as two black bands (fig. \ref{fig:profile_and_speed}.(a)).
In between these bands the light is darker than on the bright interference fringes, since the light is reflected by the oil-glass interface on the back glass plate, which is out of focus.
For a given $x$ position, by looking at the luminosity profile (fig. \ref{fig:profile_and_speed}.(b)) we can then find the center of the rivulet $z_c$.
This is done using an algorithm by fitting an heuristic idealized profile to the experimental luminosity data.
By using this approach for different frames, we are able to monitor how $z_c$ evolves with time (fig. \ref{fig:profile_and_speed}.(c)).
This allows us to measure experimentally the amplitude of the movement as well as the instantaneous speed of the rivulet.
Since the rivulet is to a good approximation of constant width, this speed corresponds to the deposition speed of the menisci which dictates the deposited film height.

\begin{figure}[b]
  \centering
  \includegraphics[scale=1]{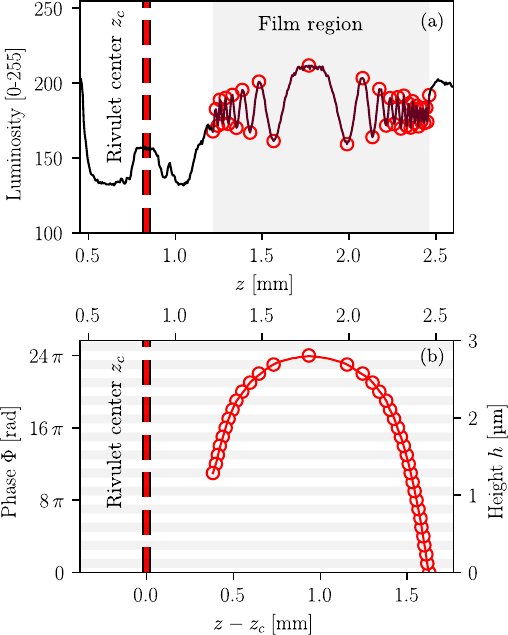}
  \caption{
    (a) Luminosity profile as a function of the transverse coordinate $z$ at $x=x_3$.
In gray, the zone of interest over which the thin film extends.
The empty circles correspond to extrema.
    (b) Phase/height profile as a function of absolute/relative transverse coordinate.}
  \label{fig:demodulation}
\end{figure}

When looking at the luminosity profile for a given $x$ coordinate we can also identify the zone over which the film extends.
This corresponds to the grayed-out zone in fig.~\ref{fig:demodulation}.(a).
We then use an algorithm to find the maxima and minima of
  luminosity, i.e. the position of the bright and dark fringes
  resulting from constructive and destructive interference.
To obtain the phase, we can associate each of the extrema to a phase shift of $\pi$ and plot
the result as a function of the transverse coordinate as shown in
figure~\ref{fig:demodulation}(b).

Last, note that it is possible to obtain information on the height in
regions where the fringes are between dark and bright, i.e. to achieve
sub-fringe resolution. To this end an analytic signal is
constructed by means of the Hilbert transform on the luminosity
signal.
From the local phase of this signal the interpolated profile is then
computed, and represented as continuous line in
fig~\ref{fig:demodulation}.(b).
This allows for a spatially well resolved measurement of the film height $h$.
Note the scale of the bottom axis for figure~\ref{fig:demodulation}.(b): we use the distance relative to the rivulet center, which allows us to make meaningful comparisons between profiles at different $x$ positions.

\begin{figure*}%[H]
  \includegraphics[scale=1]{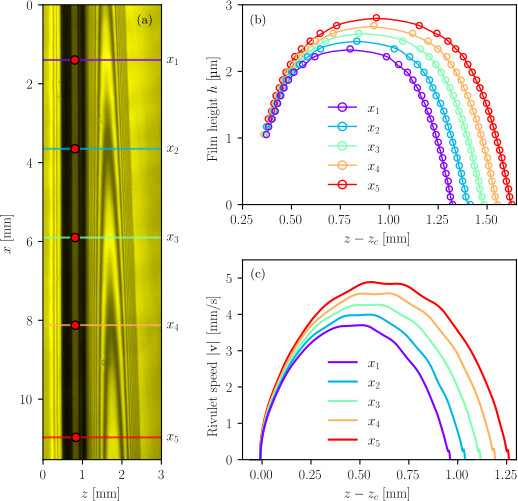}
\caption{
  (a) The reference image, with the positions of interest (colored strips).
The rivulet position $z_c$ is indicated by red circles.
  (b) The height of the film at the positions of interest as a function of the transverse coordinate relative to the rivulet position.
  (c) Speed of the rivulet as a function of relative position (same scale as (b)).}
\label{fig:compareheightandspeed}
\end{figure*}

For the experiments in this section, we use static elements in the cell that are designed to focus the flow of air generated by the speakers in a particular region.
This creates an inhomogeneous forcing, where the displacement of the rivulet depends on the $x$ coordinate.
Since the amplitude of the movement of the rivulet depends on space, so does the speed at which it moves and thus the height of the film it deposits.
An illustration of this is presented of figure~\ref{fig:compareheightandspeed}.
On the left, we see the image of reference ($t=0$) from which we extract the height profile $h(z)$ for different $x$ positions $x_{1, 2, \ldots}$.
As for figure~\ref{fig:demodulation}, the empty circles correspond to dark or bright fringes, and the full lines to the unwrapped instantaneous phases of the analytic signals obtained by Hilbert transforms.
One might remark that on fig.~\ref{fig:compareheightandspeed}.(b), the top of the profiles are one fourth of a wavelength from each other.
This is due to the fact that the selected positions $x_{1, 2, \ldots}$ are on
dark and bright interference fringes, alternatively.

Fig~\ref{fig:compareheightandspeed}.(c) is a parametric plot of the $z$-wise velocity of the
rivulet center $\text{v}(t)$ as function of its position $z(t)$ over time $-T/2 < t < 0$, with $T$ the period of the movement.
One can see by comparing fig~\ref{fig:compareheightandspeed}.(b) and (c)
that at the $x$ positions where the rivulet went faster, it deposited a higher film.
Note that since the velocity is measured for the center of the
rivulet, the two bottom axes do not correspond to one another : the
height and speed curves are shifted with respect to one another.
The shift corresponds to the distance between the center of the rivulet and the point where the film detaches from the meniscus.
We measure this shift to be of \SI{340\pm10}{\micro\meter}, which is coherent with the fact that menisci are arcs of circles of radius \SI{300}{\micro\meter}, and that the distance between the two menisci is around \SI{100}{\micro\meter}.

We now want to quantitatively make a link between the rivulet speed and the deposited film height.
This situation is analogous to the Landau--Levich-Derjaguin problem of film deposition by a static meniscus onto a moving plate, in a confined geometry where the curvature of the meniscus is not determined by gravity but by the geometry of our cell.
To characterize the speed of the moving interface we thus follow LLD and use the capillary number $\Ca = \rho\,\nu\, \abs{\dot{z_c}}/\gamma$ ; and the height of the deposited film is normalized by the curvature radius of the static meniscus $b/2$, so that we are interested in $\hnorm = h / (b/2)$.

\begin{figure}[!h]
\centering
\includegraphics[width=\figwidth]{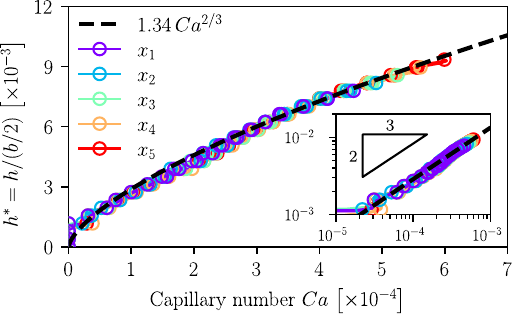}
\caption{Height of the film deposited by a periodic motion of the rivulet as a function of the capillary number. The dashed line corresponds to the LLD prediction. Inset: same data on logarithmic scale.}
\label{fig:llscaling}
\end{figure}

The height of the liquid film deposited should then be $\hnorm = 1.34\, \Ca^{2/3}$~\cite{landau1942,derjaguin1943, cantat2013}.
We plot on figure~\ref{fig:llscaling} the dimensionless height as a function of the capillary number, as well as the LLD scaling with the theoretically predicted prefactor.
We find perfect agreement between our experimental measurements and the LLD prediction, confirming our ability to generate films of deterministic height by controlling the rivulet speed.

One last point to take into consideration is the stability of these films.
In section~\ref{sec:drainage}, we studied a film generated by using a snapping motion of the rivulet.
This implied a very fast reconfiguration of the film geometry, with the film height diminishing as fast as \SI{100}{\nano\meter/\second}.
For the data presented in this section, the rivulet adopts a smooth movement with an amplitude that varies very slowly with the $x$ coordinate.
In consequence, the draining is expected to be, and indeed is, much slower.

To measure the film thinning by drainage, it can be
advantageous to correlate the luminosity signal along the crest line for
different times, rather than directly evaluating the
time derivative of the $\hcrest(x, t)$ field as done in
section~\ref{sec:drainage}.
The correlation is performed on images where the moving meniscus is
far from the crest line so that the latter evolves only through
drainage. This technique remains accurate even at very low fringe
displacement speed, of order \SI{100}{\micro\meter/s}, which
corresponds to less than half a pixel per frame.
In our test case, we estimate the height decrease at the crest to be
\SI{9\pm1}{\nano\meter/\second} --- the measurement being done over a time
of only \SI{0.2}{\second}. Being able to accurately measure such small
quantities with an excellent temporal resolution using a macroscopic
observation method is a testimony to the precision of our measurement
technique.
\section{Conclusion / perspectives} \label{sec:conclusion}
In this article, we have exposed a powerful way to deterministically create and study with precision very thin films of liquid over a rigid plate.
By injecting oil into an air-filled Hele-Shaw channel, we generate a
rivulet of liquid presenting two menisci of fixed curvature.
By pushing air into the cell, we are able to impose a transverse movement on the rivulet, and thus the displacement speed of the menisci.
The control of both the velocity of the menisci and their curvature (imposed by the cell geometry) allows us to predict the height of the film.

Using an interferometric method, we can measure the film height in the
whole region visible to the camera where the film is weakly inclined.
This measurement is instantaneous, so that the spatio-temporal
evolution of these films can be monitored with a resolution limited
only by the camera capabilities. Thanks to these accurate
measurements, we are able for example to predict the height decrease
of a film, and to attribute it to the combination of capillary and
gravitational effects.

There are of course limitations on the accuracy of our methods.
When using a rivulet confined in a Hele-Shaw cell, one must stay below a certain flow rate to avoid spontaneous meandering \cite{daerr2011}.
One must also ensure not to move the rivulet too fast with too much
amplitude, lest the rivulet be prone to a phase-locking instability
\cite{lelay2025}. Fast image acquisition necessitates a powerful
monochromatic light source.

The interferometric method imposes a trade-off between precision on
height measurement and ability to explore high-slope regions of the
film. Indeed, when using small wavelength light, the better height
resolution comes at the cost of having very thin fringes in the most
inclined parts of the film. Optically resolving such thin fringes
requires excellent spatial resolution, which is not always available.
On the contrary, using long wavelength light allows to see a greater
region of the film, at the cost of reduced precision on the height
measurement over the flatter parts.

There are also ways to improve the technique we present here, or adapt it to study other situations.
Concerning film generation, the use of a fluid rivulet confined in a cell imposes the Hele-Shaw geometry with two plates.
It would however be possible to adapt this work using a fluid rivulet
flowing on a single plate, or using sliding droplets.
To move the rivulet, we use speakers, which have a limited stroke.
Replacing the speakers by compressed air nozzles controlled by valves would allow for a greater movement amplitude and a finer spatial control of the rivulet motion.
Using a heated plate and renewing the air inside the cell would also allow for evaporation to occur,
which can be useful to cure the deposited film or enhance heat exchange in industrial applications~\cite{pagliarini2023}.%

For the purpose of open access, the author has applied
a Creative Commons Attribution (CC BY) license to any
Author Accepted Manuscript version arising from this
submission.

\providecommand{\noopsort}[1]{}\providecommand{\singleletter}[1]{#1}%

\end{document}